\documentclass[journal]{IEEEtran}

\usepackage{cite}

\usepackage[utf8]{inputenc}
\usepackage[pdftex]{graphicx}
\usepackage{float}

\usepackage{hyperref}
\usepackage{url}

\begin{document}


\title{MIMIR: Deep Regression for Automated Analysis of UK Biobank Body MRI}

\author{Taro Langner, Andr{\'e}s Mart{\'i}nez Mora, Robin Strand, H\r{a}kan Ahlstr\"{o}m, and Joel Kullberg

\thanks{Department of Surgical Sciences, Uppsala University, Sweden (T.L., AM.M. R.S., H.A., J.K.); Department of Information Technology, Uppsala University, Sweden (R.S.); and Antaros Medical AB, Sweden (H.A., J.K.).}
\thanks{\textbf{Address correspondence to} T.L. (e-mail: \textit{taro.langner@surgsci.uu.se}).}

}


\maketitle

\begin{abstract}


UK Biobank (UKB) conducts large-scale examinations of more than half a million volunteers, collecting health-related information on genetics, lifestyle, blood biochemistry, and more. Medical imaging of 100,000 subjects, with 70,000 follow-up sessions, enables measurements of organs, muscle, and body composition. With up to 170,000 mounting MR images, various methodologies are accordingly engaged in large-scale image analysis.
This work presents an experimental inference engine that can automatically predict a comprehensive profile of subject metadata from UKB neck-to-knee body MRI. It was evaluated in cross-validation for baseline characteristics such as age, height, weight, and sex, but also measurements of body composition, organ volumes, and abstract properties like grip strength, pulse rate, and type 2 diabetic status. It predicted subsequently released test data covering twelve body composition metrics with a 3\% median error.
The proposed system can automatically analyze one thousand subjects within ten minutes, providing individual confidence intervals. The underlying methodology utilizes convolutional neural networks for image-based mean-variance regression on two-dimensional representations of the MRI data. This work aims to make the proposed system available for free to researchers, who can use it to obtain fast and fully-automated estimates of 72 different measurements immediately upon release of new UKB image data.




%
\end{abstract}

\IEEEpeerreviewmaketitle

\section{Introduction}

UK Biobank (UKB) has shared an extensive record of health-related data for more than half a million volunteers with over 2,000 research projects. Starting in 2014, medical imaging was initiated for 100,000 participants with several modalities such as DXA, ultrasound, and MRI \cite{littlejohns2020uk}. Beyond collecting data on genetics, lifestyle, and biochemistry of blood and urine, ongoing research is accordingly engaged in large-scale analysis of the extensive acquired imaging data.

Information on body composition \cite{west_feasibility_2016} and liver fat content \cite{wilman2017characterisation} can be extracted from these images, with important implications for metabolic and cardiovascular disease \cite{linge2018body}. 
Although the required image analysis can involve human labor, fully-automated approaches more recently introduced neural networks for segmentation of the liver \cite{irving2017deep}, pancreas \cite{basty2020automated, bagur2020pancreas}, kidneys \cite{langner2020kidney}, and various other organs, muscles, and tissues within this study \cite{liu2020systematic, kart2021deep}. Many of these techniques target the UKB neck-to-knee body MRI, which can represent almost all of human anatomy in one comprehensive image. Lean and adipose tissue can be clearly distinguished in these images, based on a two-point Dixon technique that acquires separate, volumetric water and fat signal images \cite{west_feasibility_2016}.

Many relevant measurements nonetheless cover only a small fraction of the 40,000 images that have been available for more than a year. The evaluation of the image data, quality control, and sharing of measurements is accordingly lagging far behind the image acquisition. Researchers with access to the data, who are conducting correlation analyses to genetics, lifestyle, and blood biochemistry, but also longitudinal developments related to aging, are accordingly confined to comparably small sample sizes. Months or years can pass until measurements for newly released images become available, and this backlog can only be expected to grow over time as more images accumulate \cite{littlejohns2020uk}.

This work presents \textit{\textbf{M}edical \textbf{I}nference on \textbf{M}agnetic resonance images with \textbf{I}mage-based \textbf{R}egression} (MIMIR), an experimental software that can infer a wide range of these measurements automatically.\footnote{\url{https://github.com/tarolangner/ukb_mimir}} It can process UKB neck-to-knee body MRI, or image data from other studies that reproduce the protocol on similar demographics, and predict 72 measurements together with individual confidence intervals. In this work, the prediction model was created, retrospectively validated, and tested against subsequently released data. The value of this system consists in making automated measurements conveniently available to researchers, months or years ahead of time, almost immediately whenever new UKB images are released.

\section{Methods}

Based on convolutional neural networks, MIMIR performs an image-based, deep regression \cite{langner2021deep}. A similar methodology was independently proposed for UKB brain MRI to estimate human age and study correlations to genetics data \cite{jonsson2019brain}.

\begin{figure*}[h]
	\centering	
	\includegraphics[width=\textwidth]{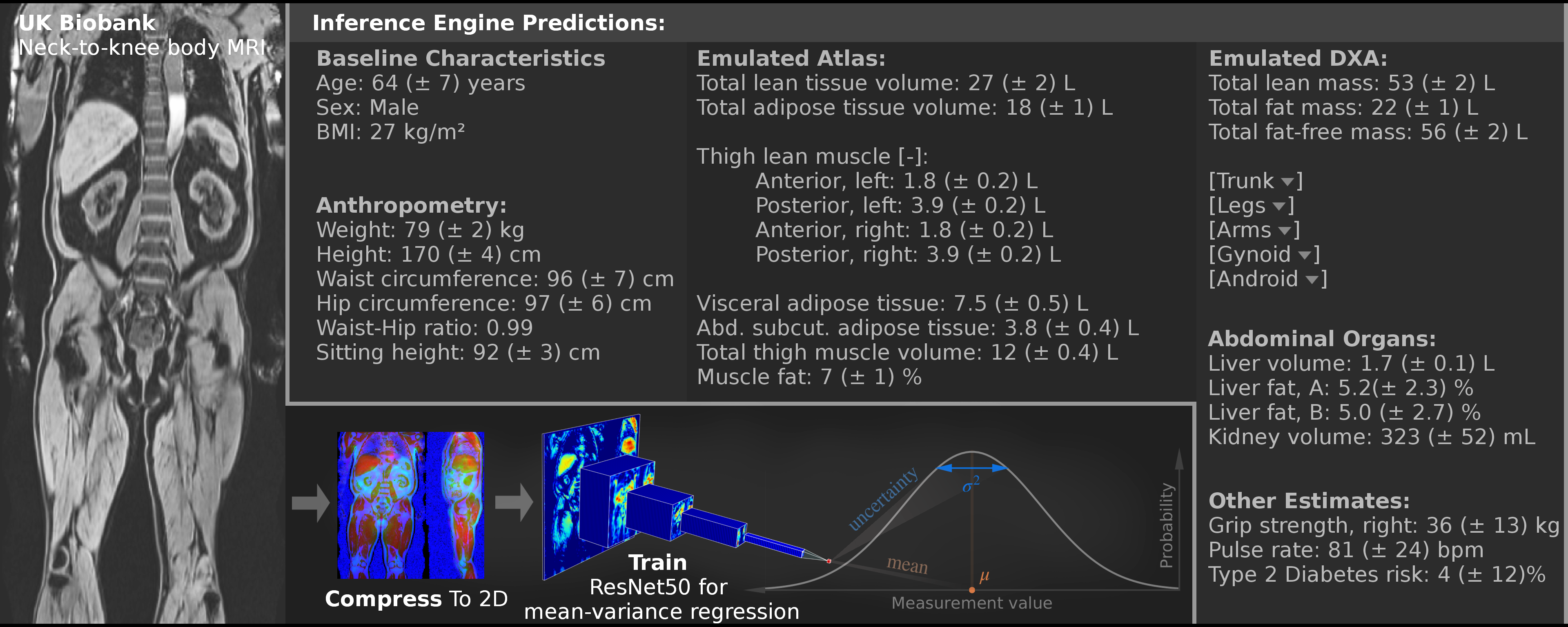}		
	\caption{MIMIR is an experimental inference engine for prediction of measurements and metadata from UKB neck-to-knee body MRI with image-based deep regression. Here, image data for one subject (left) was compressed and processed as part of the validation set. The listed measurements were inferred, together with 95\% confidence intervals based on the mean and variance predicted by a ResNet50 convolutional neural network (bottom part).}
	\label{fig_mimir}	
\end{figure*}

The neural networks process UKB neck-to-knee body MRI, compressed into a two-dimensional format with projected signals, as seen in Fig.~\ref{fig_mimir}, and predict the mean and variance of a Gaussian probability distribution over each given target measurement \cite{lakshminarayanan2017simple}. This provides both a point estimate $\mu$ for the measurement itself and a heteroscedastic variance $\sigma^2$, which can express aleatoric uncertainty and describe confidence- or prediction intervals \cite{kendall2017uncertainties}. This methodology was previously proposed for six measurements of body composition on the same data \cite{langner2021uncertainty}. Here, neural network instances were not formed into ensembles, however, in favor of speed and convenience. The targets were furthermore expanded to 72 different measurements, ranging from age, height, weight, and sex over body composition and organ volumes to experimental properties like grip strength, pulse rate, and type 2 diabetic status \cite{eastwood2016algorithms}.

UKB shared 1.5T neck-to-knee body MRI from three imaging centers for about 40,000 men and women aged 44-82 (mean 64) years, BMI 14-62 (mean 27) kg/m$^2$ and 95\% self-reported white British ethnicity (see also \cite{littlejohns2020uk, west_feasibility_2016}). This work was approved by UKB and the responsible Swedish ethics committee. Visual inspection yielded 38,916 subjects without water-fat swaps, corrupted data, implants, severe pathologies such as large tumors, or non-standard positioning \cite{langner2021uncertainty}.

From the UKB metadata, 72 fields were selected as regression targets. They were grouped into four modules, for each of which one network instance with one or more outputs was trained: Body composition, abdominal organs, anthropometric/experimental estimates, and age. The available reference values for these regression targets originate from previously shared atlas-based segmentation results \cite{west_feasibility_2016}, manual analyses \cite{wilman2017characterisation}, DXA imaging \cite{littlejohns2020uk}, and prior neural network segmentations of the liver and kidneys \cite{langner2020kidney}. Across all subjects, $73\%$ of target values were not available through UKB. Estimating these missing measurements and providing them automatically for future UKB images was the main motivation for this work.

The method was evaluated in cross-validation on a stratified, 10-fold split by grouping all subjects into ten even subsets, each of which in turn served for validation of a neural network trained on the data of all remaining sets. This pretrained ResNet50 \cite{he_deep_2016} with a specialized mean-variance loss \cite{kendall2017uncertainties} is visualized in Fig.~\ref{fig_mimir}. All samples with at least one ground truth value were used by setting the loss for missing values to zero. The network was trained in PyTorch with the Adam optimizer, batch size 32, and augmentation by random translations. After training for 8,000 iterations, the learning rate was reduced from $5e-5$ to $5e-6$ for another 2,000 iterations. 

Each module was then trained on all samples to make predictions for the entire imaged cohort.
These were tested against a subsequent release of UKB reference values, covering twelve body composition targets for 15,000 subjects \cite{west_feasibility_2016}.

The predictions were evaluated with the intraclass correlation coefficient (ICC) using a two-way random, single measures, absolute agreement definition. For this metric, the reliability can be considered good for values above $0.75$ and excellent for those above $0.90$, with a maximum of $1.0$ \cite{koo2016guideline}. Additionally, the coefficient of determination (R$^2$), mean absolute error (MAE), mean absolute percentage error (MAPE) and the area under curve of the receiver operating characteristic curve (AUC-ROC) are provided.
The uncertainty estimates were calibrated with post-hoc scaling factors \cite{langner2021uncertainty}.

\section{Results and Discussion}

\begin{figure*}[h]
	\centering	
	\includegraphics[width=\textwidth]{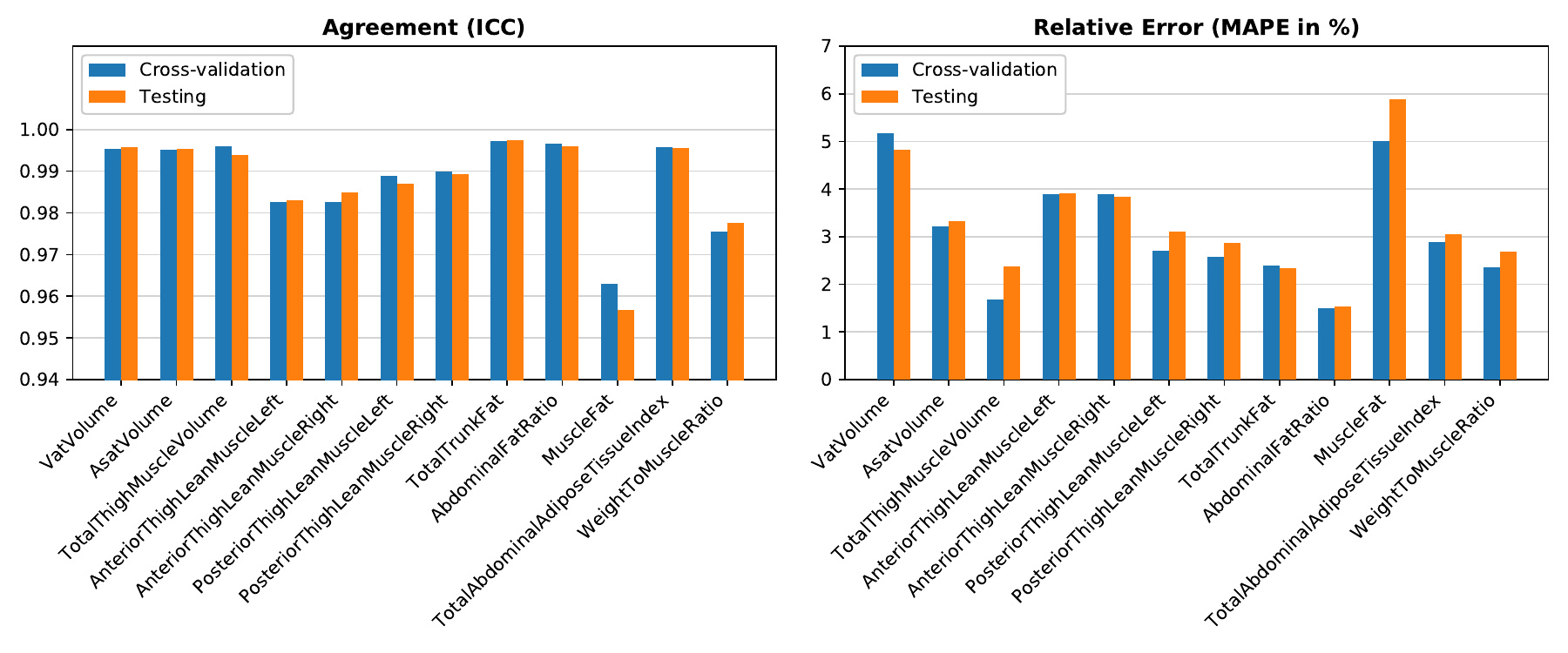}		
	\caption{The predictive performance in cross-validation closely matched the results on available test data for 12 body composition targets \cite{west_feasibility_2016} of 15,000 subjects. 
		}
	\label{fig_testing}	
\end{figure*}

\begin{figure}
	\centering	
	\includegraphics[width=\columnwidth]{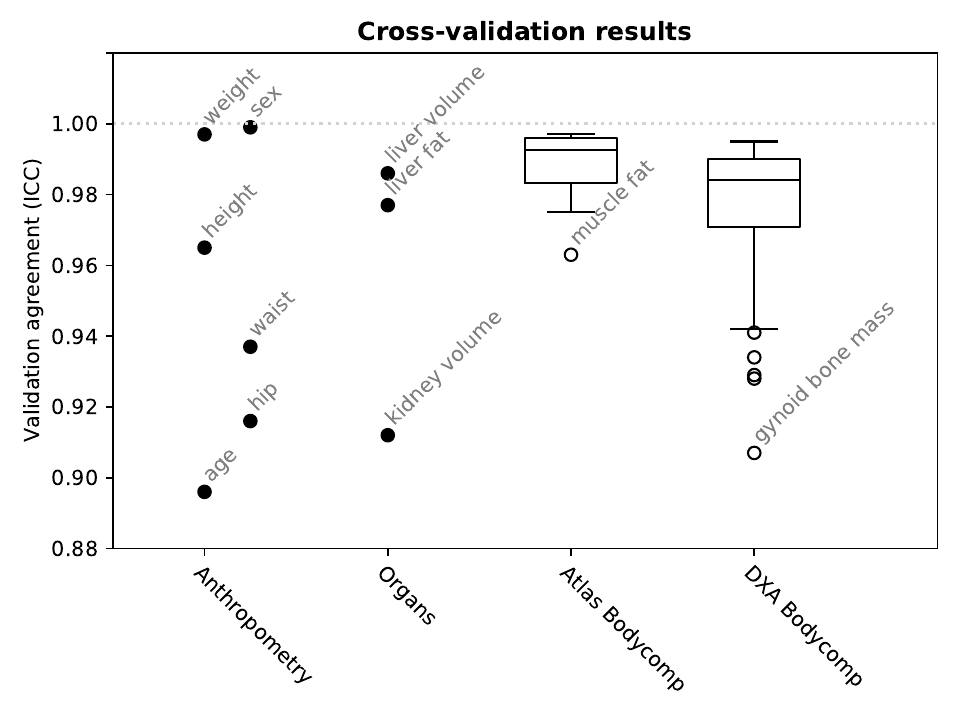}		
	\caption{Agreement between predictions and reference in cross-validation, expressed as intraclass correlation coefficient (ICC). Note that not all targets fall within the bounds of this plot.}
	\label{fig_boxplot}	
\end{figure}

Results for the test set and cross-validation are shown in Fig.~\ref{fig_testing} and \ref{fig_boxplot}. Extensive supplementary documentation for all 72 targets is available online.\footnote{\url{https://github.com/tarolangner/ukb_mimir}}
The validation yielded mean absolute errors (MAE) of 2.6 years for chronological age, 1.8 cm for height, 0.9 kg for bodyweight, and correct identification of sex in all but \mbox{14 of 38,874} subjects (of whom five differed in registered vs genetic sex). The relative error (MAPE) was below 7\% for combined parenchymal kidney volume, 3\% for liver volume, and median 3\% for twelve body composition measurements.
Liver fat was predicted with R$^2 = 0.955$. The test results closely matched the cross-validation metrics, indicating robust generalization to future UKB releases. The uncertainty estimates were likewise mostly well-calibrated in testing, so that the confidence intervals provided along with each measurement reliably represent the prediction error. The inference engine processed one thousand subjects within ten minutes on an Nvidia RTX 2080 Ti with 11GB.

Several limitations apply. The trained networks should not be expected to correctly process arbitrary MRI data. Generalization is likely restricted to the given UKB imaging protocol, device type, and demographics, whereas out-of-domain tasks may require hundreds of subjects for renewed training \cite{langner2021deep}. 

Despite identifying probable type 2 diabetics \cite{eastwood2016algorithms} with an AUC-ROC of 0.866, similar to prior work on MRI-based neural network diagnostics \cite{wagner2020102}, the corresponding $435 / 996$ $(44\%)$ precision and $435 / 1741$ $(25\%)$ recall are not viable for clinical application.

Not all metadata can be inferred with the proposed approach, as the required information may be lost by preprocessing or absent from the MRI data. 
Other measurements are clinically trivial, such as age, sex, and weight. For some, alternative techniques may be superior, such as liver fat measurements from dedicated UKB liver MRI which encodes fat fractions more faithfully. The projected, two-dimensional input format also imposes limitations on smaller structures, with the proposed kidney volume measurements almost doubling the error achieved by segmentation of axial slices \cite{langner2020kidney}. 

Whereas prior work trained separate network instances for each target \cite{Langner2020}, the proposed method can predict more information with only four multi-output networks. It also automatically indicates potential outliers and failure cases by providing confidence intervals for each predicted value, based on uncertainties that are well-calibrated even without the complexity and runtime costs of ensembling \cite{langner2021uncertainty}.

Future work may explore fully volumetric processing and other imaging protocols, which could leverage more, potentially critical information. However, several metrics can already be accurately inferred in the proposed way, such as liver volume and body composition metrics. With additional metadata becoming available, such as hormone levels, disease outcomes, and mortality statistics, more modules could be added in the future. The networks trained in this work may provide benefit for independent studies that replicate the used imaging protocol and hold potential for transfer learning.

This work proposes an implementation that can conveniently provide image-derived phenotypes at large scale for prototyping in correlation studies, months or years before the reference measurements will be available. Over the coming years, it will stand ready for fully-automated analysis of more than 100,000 upcoming UKB images that are yet to be released.

\section{Acknowledgments}

This research was enabled by grants from the Swedish Heart-Lung Foundation and the Swedish Research Council (2016-01040, 2019-04756, 2020-0500, 2021-70492) and the UK Biobank Resource under application no. 14237.

\bibliographystyle{ieeetr}
\bibliography{references}

\begin{thebibliography}{10}

\bibitem{littlejohns2020uk}
T.~J. Littlejohns, J.~Holliday, L.~M. Gibson, S.~Garratt, N.~Oesingmann,
  F.~Alfaro-Almagro, J.~D. Bell, C.~Boultwood, R.~Collins, M.~C. Conroy, {\em
  et~al.}, ``The uk biobank imaging enhancement of 100,000 participants:
  rationale, data collection, management and future directions,'' {\em Nature
  Communications}, vol.~11, no.~1, pp.~1--12, 2020.

\bibitem{west_feasibility_2016}
J.~West, O.~Dahlqvist~Leinhard, T.~Romu, R.~Collins, S.~Garratt, J.~D. Bell,
  M.~Borga, and L.~Thomas, ``Feasibility of mr-based body composition analysis
  in large scale population studies,'' {\em PloS one}, vol.~11, no.~9,
  p.~e0163332, 2016.

\bibitem{wilman2017characterisation}
H.~R. Wilman, M.~Kelly, S.~Garratt, P.~M. Matthews, M.~Milanesi, A.~Herlihy,
  M.~Gyngell, S.~Neubauer, J.~D. Bell, R.~Banerjee, {\em et~al.},
  ``Characterisation of liver fat in the uk biobank cohort,'' {\em PloS one},
  vol.~12, no.~2, p.~e0172921, 2017.

\bibitem{linge2018body}
J.~Linge, M.~Borga, J.~West, T.~Tuthill, M.~R. Miller, A.~Dumitriu, E.~L.
  Thomas, T.~Romu, P.~Tun{\'o}n, J.~D. Bell, {\em et~al.}, ``Body composition
  profiling in the uk biobank imaging study,'' {\em Obesity}, vol.~26, no.~11,
  pp.~1785--1795, 2018.

\bibitem{irving2017deep}
B.~Irving, C.~Hutton, A.~Dennis, S.~Vikal, M.~Mavar, M.~Kelly, and J.~M. Brady,
  ``Deep quantitative liver segmentation and vessel exclusion to assist in
  liver assessment,'' in {\em Annual Conference on Medical Image Understanding
  and Analysis}, pp.~663--673, Springer, 2017.

\bibitem{basty2020automated}
N.~Basty, Y.~Liu, M.~Cule, E.~L. Thomas, J.~D. Bell, and B.~Whitcher,
  ``Automated measurement of pancreatic fat and iron concentration using
  multi-echo and t1-weighted mri data,'' in {\em 2020 IEEE 17th International
  Symposium on Biomedical Imaging (ISBI)}, pp.~345--348, IEEE, 2020.

\bibitem{bagur2020pancreas}
A.~T. Bagur, G.~Ridgway, J.~McGonigle, M.~Brady, and D.~Bulte, ``Pancreas
  segmentation-derived biomarkers: Volume and shape metrics in the uk biobank
  imaging study,'' in {\em Annual Conference on Medical Image Understanding and
  Analysis}, pp.~131--142, Springer, 2020.

\bibitem{langner2020kidney}
T.~Langner, A.~{\"O}stling, L.~Maldonis, A.~Karlsson, D.~Olmo, D.~Lindgren,
  A.~Wallin, L.~Lundin, R.~Strand, H.~Ahlstr{\"o}m, {\em et~al.}, ``Kidney
  segmentation in neck-to-knee body mri of 40,000 uk biobank participants,''
  {\em Scientific reports}, vol.~10, no.~1, pp.~1--10, 2020.

\bibitem{liu2020systematic}
Y.~Liu, N.~Basty, B.~Whitcher, J.~Bell, E.~Sorokin, N.~van Bruggen, E.~L.
  Thomas, and M.~Cule, ``Genetic architecture of 11 organ traits derived from
  abdominal mri using deep learning,'' {\em ELife}, p.~10:e65554, 2021.

\bibitem{kart2021deep}
T.~Kart, M.~Fischer, T.~K{\"u}stner, T.~Hepp, F.~Bamberg, S.~Winzeck,
  B.~Glocker, D.~Rueckert, and S.~Gatidis, ``Deep learning--based automated
  abdominal organ segmentation in the uk biobank and german national cohort
  magnetic resonance imaging studies,'' {\em Investigative Radiology}, 2021.

\bibitem{langner2021deep}
T.~Langner, R.~Strand, H.~Ahlstr{\"o}m, and J.~Kullberg, ``Deep regression for
  uncertainty-aware and interpretable analysis of large-scale body mri,'' 2021.

\bibitem{jonsson2019brain}
B.~A. J{\'o}nsson, G.~Bjornsdottir, T.~Thorgeirsson, L.~M. Ellingsen, G.~B.
  Walters, D.~Gudbjartsson, H.~Stefansson, K.~Stefansson, and M.~Ulfarsson,
  ``Brain age prediction using deep learning uncovers associated sequence
  variants,'' {\em Nature communications}, vol.~10, no.~1, pp.~1--10, 2019.

\bibitem{lakshminarayanan2017simple}
B.~Lakshminarayanan, A.~Pritzel, and C.~Blundell, ``Simple and scalable
  predictive uncertainty estimation using deep ensembles,'' in {\em Advances in
  neural information processing systems}, pp.~6402--6413, 2017.

\bibitem{kendall2017uncertainties}
A.~Kendall and Y.~Gal, ``What uncertainties do we need in bayesian deep
  learning for computer vision?,'' in {\em Advances in neural information
  processing systems}, pp.~5574--5584, 2017.

\bibitem{langner2021uncertainty}
T.~Langner, F.~K. Gustafsson, B.~Avelin, R.~Strand, H.~Ahlstr{\"o}m, and
  J.~Kullberg, ``Uncertainty-aware body composition analysis with deep
  regression ensembles on uk biobank mri,'' {\em arXiv preprint
  arXiv:2101.06963}, 2021.

\bibitem{eastwood2016algorithms}
S.~V. Eastwood, R.~Mathur, M.~Atkinson, S.~Brophy, C.~Sudlow, R.~Flaig,
  S.~de~Lusignan, N.~Allen, and N.~Chaturvedi, ``Algorithms for the capture and
  adjudication of prevalent and incident diabetes in uk biobank,'' {\em PloS
  one}, vol.~11, no.~9, p.~e0162388, 2016.

\bibitem{he_deep_2016}
K.~He, X.~Zhang, S.~Ren, and J.~Sun, ``Deep {Residual} {Learning} for {Image}
  {Recognition},'' in {\em 2016 {IEEE} {Conference} on {Computer} {Vision} and
  {Pattern} {Recognition} ({CVPR})}, pp.~770--778, June 2016.

\bibitem{koo2016guideline}
T.~K. Koo and M.~Y. Li, ``A guideline of selecting and reporting intraclass
  correlation coefficients for reliability research,'' {\em Journal of
  chiropractic medicine}, vol.~15, no.~2, pp.~155--163, 2016.

\bibitem{wagner2020102}
R.~Wagner, B.~Dietz, J.~Machann, P.~Schwab, J.~K. Dienes, S.~Reichert, A.~L.
  Birkenfeld, H.-U. Haering, F.~Schick, N.~Stefan, {\em et~al.}, ``102-or:
  Detection of diabetes from whole-body magnetic resonance imaging using deep
  learning,'' 2020.

\bibitem{Langner2020}
T.~Langner, R.~Strand, H.~Ahlstr{\"o}m, and J.~Kullberg, ``Large-scale biometry
  with interpretable neural network regression on uk biobank body mri,'' {\em
  Scientific reports}, vol.~10, no.~1, pp.~1--9, 2020.

\end{thebibliography}

\end{document}